\documentclass[12pt]{article}
\usepackage{amsmath,amsfonts,amssymb,latexsym,stmaryrd,mathrsfs}

\def\be{\begin{equation}}
\def\ee{\end{equation}}
\def\bq{\begin{eqnarray}}
\def\eq{\end{eqnarray}}
\def\beq{\begin{eqnarray*}}
\def\eeq{\end{eqnarray*}}

\begin{document}
\begin{titlepage}

\begin{center}
{\Huge Topology of the ambient boundary and the convergence of causal curves}

\vspace{1cm}
{\large Ignatios Antoniadis$^{1,2}$, Spiros Cotsakis$^{3,4\ddagger}$ }\\

\vspace{0.5cm}

$^1$ {\normalsize Albert Einstein Center, Institute for Theoretical Physics}\\
{\normalsize Bern University, Sidlerstrasse 5 CH-3012 Bern, Switzerland}\\

$^2$ {\normalsize LPTHE, UMR CNRS 7589}\\ {\normalsize Sorbonne Universit\'es, UPMC Paris 6, 75005 Paris, France}\\

$^3$ {\normalsize School of Applied Mathematics and Physical Sciences \\
National Technical University, 15780 Athens, Greece} \\

\vspace{2mm} {\normalsize {\em E-mails:}
$^\dagger$\texttt{antoniad@lpthe.jussieu.fr},
$^\ddagger$\texttt{skot@aegean.gr}

\vspace{0.5cm}

\today}
\end{center}

\begin{abstract}
\noindent We discuss the topological nature of the  boundary spacetime, the conformal infinity of the ambient cosmological metric. Due to the existence of a homothetic group, the bounding spacetime must be equipped not with the usual Euclidean metric topology but with the Zeeman fine topology. This then places severe constraints to the convergence of a sequence of causal curves and the extraction of a limit curve, and also to our ability to formulate conditions for singularity formation.

\end{abstract}

\begin{center}
{\line(5,0){280}}
\end{center}

%\noindent$^3${\small Also Ecole Polytechnique, F-91128 Palaiseau, France}

\noindent$^4${\small On leave from the University of the Aegean, 83200 Samos, Greece.}

\end{titlepage}
%%%%%%%%%%%%%%%%%%%%%%%%%%%%%%%%%%%%%%%%%%%%%%%%%%
%%%%%%%%%%%%%%%%%%%%%%%%%%%%%%%%%%%%%%%%%%%%%%%%%%
%%%%%%%%%%%%%%%%%%%%%%%%%%%%%%%%%%%%%%%%%%%%%%%%%%

\section{Introduction}
\noindent For half a century it has been known  that  general relativity has the remarkable property of generically leading to spacetime singularities where it must break down as a physical theory and together our ability to further predict the future or past of spacetime come to a halt \cite{pen65,hp}. The mathematical results that lead to the prediction of spacetime singularities in general relativity, the so-called singularity theorems, rest on the import of topological methods in relativity, in particular, global methods of causality theory in Lorentzian manifolds and  the behaviour of geodesics in spacetime, and show that under various physically plausible and geometric conditions, spacetime must be geodesically incomplete \cite{pen72,he,o'neill,wald}.

Hence, general relativity as a physical theory predicts that there must be a collapse singularity in the past of the universe's history, known as the big bang, and also in the future, localized `inside' black holes (assuming cosmic censorship), or in the form of a naked singularity. It is the existence of spacetime singularities as it follows from the singularity theorems that is to a large measure responsible for the long-held view that quantum gravitational effects may be directly required also  for a restoration of the loss of classical predictability in the gravitational collapse to a black hole and in various asymptotic questions of modern cosmology, in much the same way as in the `quantum rescue' of classical newtonian evolution (cf. e.g., \cite{pen04}, various chapters).

Recently we have examined the question of spacetime singularities in general relativity by introducing the notion of an ambient cosmological metric. In the resulting `ambient cosmology', a number of questions such as the existence of spacetime singularities and cosmic censorship acquire new meaning and find novel interpretations. In this paper, we discuss a consequence of the existence of a new principle in the boundary spacetime of the ambient cosmological metric. This is related to the topological nature of the boundary and leads to an impossibility of having converging families of causal curves.

The plan of this paper is as follows. In the next section we review basic aspects of the ambient construction including the basic asymptotic condition valid on the spacetime 4-boundary $M$. In Section 3, we introduce the Zeeman topology and point out that it is that which is uniquely adapted to the boundary spacetime in the ambient construction. In Section 4, we show that due to the topological structure of the boundary, it is not possible to have convergent families of causal curves  on $M$ and therefore the existence of a limit curve is lost in the bounding spacetime equipped with the Zeeman topology. In the last Section, we discuss these results especially their connection with the impossibility of formulating singularity theorems for the conformal infinity of the ambient cosmological metric.

\section{Ambient cosmology}
The basic idea of ambient cosmology \cite{ac} is to regard our 4-dimensional spacetime as a kind of bounding hypersurface, the conformal infinity,  of a new cosmological metric in 5-dimensional `ambient' space. Although the ambient metric is a well-known construction in conformal geometry \cite{fef}, however, in our approach we use the word `ambient' with a different meaning, and this leads to a completely new condition imposed on the boundary, instead of the usual uniqueness of the ambient 5-metric corresponding to two conformally related 4-metrics on the boundary.

Let us briefly explain some of the required ideas,  referring  to \cite{ac} for complete proofs. We start with a 4-dimensional spacetime $M$ on which a conformal structure $[g_4]$ is defined, that is we have a family of conformally related 4-metrics on $M$, much like as in the conformal method, cf. \cite{pr}, chap. 9. Our ambient metric $g_+$ is then defined at each point $p=(x^\mu,w)$ of the `ambient' space $V=M\times \mathbb{R}$, where $\mathbb{R}$ parametrized by $w$ denotes the fifth dimension, and can be brought in the normal form $g_+=w^{-n}\left(\sigma^2(w)g_4(x^\mu)+dw^2\right), n\in\mathbb{Q}^+$, as $w\rightarrow 0$, with $\sigma(w)$ a smooth (infinitely differentiable) function such that $\sigma(0)$ is a nonzero constant.

In this asymptotic sense, the first factor in the cartesian product, $M$, is located at the limiting position $w=0$ (see also below), and everything we do may be described as `expanding around $w=0$'.
The metric $g_+$ solves the 5-dim Einstein equations $G_{AB}=T_{AB}$ on $V$,
where $A,B=0,1,2,3,4$, and $T_{AB}$ is the stress tensor of an analog of a perfect fluid with equation of state $P=\gamma \rho$, where the `pressure' $P$ and the `energy density' $\rho$ depend only on the extra spacelike dimension $w$, and the density $\rho_+(w)$ has the generic form $\rho_+=w^{-z}\theta(w)$, $z\in\mathbb{Z}$ and $\theta(w)$ smooth.

Our ambient  metric $g_+$ has two further properties of special significance in the present context. Firstly, $g_+$ on $V$ has $M$ as its conformal infinity equipped with the manifestly regular, non-degenerate metric at $w=0$, denoted by $\mathring{g}|_M(0)$, where $\mathring{g}=w^{n}g_+=\sigma^2 g_4+dw^2$. In this sense the bounding spacetime $M$ has become the conformal infinity of the ambient space $V$, with a 4-metric given by $\mathring{g}|_M= \sigma^2(w) g_4$.

The second property of the ambient metric $g_+$ is what we call `the asymptotic condition', and it basically reveals the presence of a homothetic symmetry on the boundary $M$. It may be described as follows (cf. \cite{ac}, Section 6 for a proof). Consider two conformally related 4-metrics $g_1,g_2$ in the conformal geometry of $M$, $g_1$ being the `bad' (roughly meaning `singular') and $g_2$ the `good' metric on the boundary (these are technical terms defined precisely in \cite{ac}). Then their ambient metrics $\mathring{g}_1|_M,\mathring{g}_2|_M$, although not equal as in the Fefferman-Graham construction \cite{fef}, differ by a homothetic transformation, that is
\be
\mathring{g}_2|_M(0)=c\,\mathring{g}_1|_M(0),\quad c:\;\textrm{const.}
\ee
Therefore, starting from a conformal geometry on the spacetime $M$, the ambient cosmological metric returns a 4-geometry on  $M$ (its conformal infinity) that has a homothetic symmetry. In \cite{ac}, we have given conclusive evidence that because of these two properties, namely, the fact that $M$ is the conformal infinity of $g_+$ and that  it also possesses a homothetic symmetry, no spacetime singularities can be developed on $M$, that is $(M,\mathring{g}|_M)$ must be free of cosmological and naked singularities, hence providing a major improvement over the standard situation in general relativity.

However, this raises the intriguing question of whether or not one may still  formulate for $M$ the conditions of the standard singularity theorems of general relativity. The purpose of the remaining of this paper is to give a negative answer to this question. The arguments advanced below will also help us  understand why one was originally able to formulate the conditions necessary for the validity of the singularity theorems as is usually done in general relativity.

\section{Uniqueness of the Zeeman topology}
Since the mathematical techniques involved in global causality theory and the singularity theorems are topological in nature, the first question we have to face is how to connect the homothetic symmetry on the conformal boundary $M$ (as a result of the ambient cosmological 5-metric on $V$) with its topological properties. There is a very neat answer to this question, provided long ago by a remarkable mathematical theorem of C. Zeeman. In 1967 he published a paper \cite{z} in which  showed that for Minkowski spacetime the group of homothetic symmetries (that is Lorentz transformations with dilatations) coincides with the group of all homeomorphisms of $M$ provided that its topology is not the usual Euclidean metric topology (that is $M$ is locally Euclidean) but a new one, called  the fine topology $\mathcal{Z}$.

The Zeeman fine topology $\mathcal{Z}$ being strictly finer than the Euclidean metric topology (it is actually the finest topology that can be defined on $M$, finest meaning the one having the most open sets), possesses various improved properties over the latter which mirror precisely the spacetime nature of $M$, rather than importing on $M$ properties from the Euclidean space $\mathbb{R}^4$ unphysical for a spacetime.

To describe $\mathcal{Z}$, we say that for $x\in M$, an open ball in $\mathcal{Z}$ has the form $B_{\mathcal{Z}}(x;r)= (B_{\mathcal{E}}(x;r)\setminus N(x))\cup {\{x\}}$, where $B_{\mathcal{E}}(x;r)$ is the Euclidean-open ball, and $N(x)$ the null cone at $x$ (we remove  $N(x)$ and put back only the point $x$). Then $B_{\mathcal{Z}}(x;r)$ is $\mathcal{Z}$-open, but not $\mathcal{E}$-open. Hence, a set $A\subset M$ is $\mathcal{Z}$-open if $A\cap B$ as a subset of $B$ is $\mathcal{E}$-open, for every spacelike plane and timelike line $B$.

Zeeman also conjectured in \cite{z} that an extension to the curved spacetimes of general relativity should be possible, that is for a general spacetime the homothetic group must be isomorphic to the homeomorphism group of the Zeeman topology, a conjecture  that  was shown to be correct by G\"{o}bel \cite{g} (see also \cite{hkm}). It was also shown in those references that amongst all possible generalized topologies, the Zeeman topology is the unique one having this property (all others have homeomorphism groups isomorphic to the conformal group). Hence,  for any spacetime $M$ in general relativity we have the freedom to choose either the standard Euclidean metric topology, giving $M$ the usual  manifold topology, or the Zeeman topology. It is of course the former that is used in all standard discussions of relativity.

%Result
However, for our bounding spacetime $M$, the conformal infinity of the ambient space $V$, it appears we do not have this freedom. Since we have shown the existence of a homothetic symmetry on $M$, it follows from the above discussion that it would be quite wrong to assume the usual  Euclidean metric topology for $M$, and we  have no choice but to think of $M$ equipped with the Zeeman topology $\mathcal{Z}$. (This is quite apart from the well-known arguments of Refs. \cite{z,g,hkm} that it would be wrong to equip a spacetime with the Euclidean rather than the Zeeman (or a similar) topology in the first place!)

\section{Impossibility of singularities}
Is there then any obstruction to the validity  of the singularity theorems in a spacetime $M$ using the Zeeman topology instead of the Euclidean metric one? One would expect that the answer would be `no' because the singularity theorems are results  that are known to be only valid for spacetimes (manifolds with a Lorentzian signature) and are certainly not true in Riemannian geometry (positive signature).

However, it is a very surprising result  that in fact the answer to the above question is that it is impossible to formulate not only the singularity theorems but also the most basic causality results for $M$ in the Zeeman topology. One notion that plays a key role in many such theorems is the convergence of a sequence of causal curves. Looking carefully at the proofs of various such results, we note that for a sequence of causal curves to converge to a limit curve one uses in an essential way the Euclidean balls with their Euclidean metric and their compactness to extract the necessary limits (cf. e.g., \cite{wald}, Prop. 8.1.5 and Theorem 9.5.2,  \cite{he}, pp. 266-71, etc).

It follows that since the Zeeman topology is strictly finer, such sequences will be Zeno sequences, that is their convergence in the Euclidean topology will not guarantee the existence of a limit curve in the Zeeman topology. The impossibility of a limit curve in this sense is very important, for it forbids the formation of a  basic contradiction present in the proofs of all singularity theorems.

 In any result of this type, a contradiction appears when assuming the existence of a curve of length greater than some maximum starting from a spacelike Cauchy surface  $\Sigma$ (on which the mean curvature is negative) downwards to the past. In any typical  proof of the singularity theorems (such as  e.g., Theorem 9.5.1 in \cite{wald}), one uses the process of extraction of a limit curve $\gamma$ (which locally maximizes the length between $\Sigma$ and an event $p$) to arrive at a contradiction in the sense that no curve can have length greater than that of the geodesic $\gamma$. If $\gamma$ cannot be extracted as a suitable limit of a converging family of causal curves on $M$, the nonexistence of  curves with  length greater than the `maximum' length of the locally maximizing geodesic (without conjugate points) $\gamma$ cannot be effected. Therefore one may not speak of incomplete geodesics on the boundary spacetime $M$ of the present context.

Hence, it follows that it may not be possible to formulate the same sufficient conditions for geodesic incompleteness for the Zeeman topology for $M$ as it is usually done using the manifold topology.

\section{Discussion}
In this paper we have used the topological nature of the boundary spacetime of the ambient cosmological 5-metric to examine the question of whether or not one can form sufficient conditions for singularity formation using convergent families of causal curves.

We reviewed the ambient cosmological construction which implies an asymptotic restriction, the existence of a homothetic symmetry on the boundary, the conformal infinity of the ambien metric. This in turn implies that the topology of the boundary must necessarily be the Zeeman fine topology instead of the standard Euclidean one.

Armed with the topological information of the boundary spacetime, we proceeded to analyze the question of singularities through the focusing of families of causal curves. We found that it is not possible to extract a limit curve from any family of causal curves of $M$. This implies that, given the homothetic symmetry on $M$, one cannot arrive at a contradiction by constructing curves of larger than some maximum length as in the standard singularity theorems.

The arguments advanced in this paper provide another proof of the impossibility of spacetime singularities on the boundary spacetime surrounding the ambient 5-space on which the Einstein equations with fluid sources hold.

\end{document}